\documentclass[final,1p,times]{elsarticle} % do not change
\usepackage{graphicx} % do not change
\usepackage{amssymb} % do not change
\usepackage{amsthm} % do not change
\usepackage{lineno} % do not change

\def\y{{\bf y}}
\def\x{{\bf x}}
\def\p{{\bf p}}
\newcommand{\as}{\alpha_{\mathrm{s}}}

\newcommand{\pt}{{\p_\perp}}

\newcommand{\ud}{\, \mathrm{d}}
\newcommand{\nc}{{N_\mathrm{c}}}

\newcommand{\qs}{Q_\mathrm{S}}

\journal{Nuclear Physics A} % do not change
\begin{document} % do not change

\begin{frontmatter} % do not change

%% QM09Author: please enter your  
%% Title, author and address info here; please do not use footnotes

% Your Title - please modify
\title{Long range rapidity correlations and the ridge in A+A collisions}

% Principle author, and co-authors - please modify
\author{Fran\c cois Gelis$^{a}$}
\author{Tuomas Lappi$^{a,b}$}
\author{Raju Venugopalan$^{c}$}

\address[a]{Institut de Physique Th\'eorique, %
B\^at. 774, CEA/DSM/Saclay, 91191 Gif-sur-Yvette, France}
\address[b]{Department of Physics, %
 P.O. Box 35, 40014 University of Jyv\"askyl\"a, Finland}
 \address[c]{Physics Department, Brookhaven
National Laboratory, Upton, NY 11973, USA}

\begin{abstract} % do not change
%% Text of abstract goes here - please modify
We discuss results for  n-gluon correlations that form the basis of the Glasma flux tube picture of early times in heavy ion collisions. Our formalism 
 is valid to all orders in perturbation theory at leading logarithmic accuracy in $x$ and includes  both QCD bremsstrahlung and the many body screening and recombination effects 
 that are important at  large parton densities.  Long range rapidity correlations, as seen in the near-side ridge in heavy ion collisions, are a chronometer of these early time strong 
color field dynamics. They also contain information on how radial flow develops in heavy ion collisions. 

\end{abstract} % do not change

\end{frontmatter} % do not change

%% QM09: we keep linenumbers at least for initial version
%\linenumbers % do not change

%% start of main text - please modify. Below is a sub-set (single section) 
%% of an earlier proceedings that show how one can handle references 
%% and figures etc.
%%\section{}\label{}

Recently, high energy factorization theorems were derived~\cite{Gelis:2008rw}  for inclusive multi-gluon 
production in a rapidity interval $\Delta y \lesssim 1/\as$
in A+A collisions.  This work is based on a field theory formalism for particle production from strong time dependent sources~\cite{Gelis:2006yv}. Our result can be expressed very compactly as~\cite{Gelis:2008rw}
%\begin{widetext}
\begin{eqnarray}
\left<\frac{\ud^n  N_n}{\ud^3\p_1\cdots \ud^3\p_n}\right>_{_{\rm LLog}}
 =
  \int \big[D\rho_1\big]\big[D\rho_2\big]\;
  Z_{y}\big[\rho_1\big]\,
  Z_{y}\big[\rho_2\big]\;
  \left.\frac{\ud N}{\ud^3\p_1}\right|_{_{\rm LO}}\cdots
  \left.\frac{\ud N}{\ud^3\p_n}\right|_{_{\rm LO}}\; .
\label{eq:ngluon-LLog}
\end{eqnarray}
%\end{widetext}
In the Color Glass Condensate formalism~\cite{McLerran:1994ni}, the $Z$'s are gauge invariant weight functionals that describe the distribution of color
 sources in each nucleus  at the rapidity of interest and  are obtained by evolving the 
JIMWLK equations~\cite{JalilianMarian:1998cb} from an initial rapidity close to the 
beam rapidity. The JIMWLK equations contain information about multi-parton correlations at all rapidities in the nuclear wavefunctions. In the large $\nc$ limit, for large nuclei, the $Z$'s
can be obtained from the simpler mean field BK~\cite{Balitsky:1998kc} equation.

The leading order single particle distributions in 
Eq.~(\ref{eq:ngluon-LLog}) are given by 
\begin{eqnarray}
\left.E_\p{\frac{\ud N}{\ud^3\p}}\right|_{_{\rm LO}} \Big[\rho_1,\rho_2\Big]=\frac{1}{16\pi^3}
\lim_{x_0\to+\infty}\int d^3\x \,d^3\y
\;e^{ip\cdot(x-y)}
\;(\partial_x^0-iE_\p)(\partial_y^0+iE_\p)
\nonumber \\
\times\sum_{\lambda}
\epsilon_\lambda^\mu(\p)\epsilon_\lambda^\nu(\p)\;
A_\mu^{\rm cl.}[\rho_1,\rho_2](x)\;A_\nu^{\rm cl.}[\rho_1,\rho_2](y)\; .
\label{eq:AA}
\end{eqnarray}
For each configuration of sources $\rho_{1,2}$ of each nucleus, 
one solves classical Yang--Mills equations to compute the gauge fields $A_\mu^{\rm cl.}[\rho_1,\rho_2]$ in the forward light cone~\cite{Krasnitz:2003jw}. 
Equation~(\ref{eq:AA}), when substituted 
in Equation~(\ref{eq:ngluon-LLog}), gives, to all orders in perturbation theory and to leading logarithmic accuracy in $x$, the n-gluon inclusive distribution in high energy A+A collisions. 
Eq.~(\ref{eq:ngluon-LLog}) is valid when all the produced particles are measured 
within a rapidity interval $ \lesssim 1/\as$ from each other;  $Z$ is evaluated for rapidity $y_1 \approx \cdots \approx y_n \approx y$. 
The expression in Eq.~(\ref{eq:ngluon-LLog}) suggests that in a single event--corresponding to a particular configuration of color sources--the leading contribution is from $n$ tagged gluons that are produced independently. The correlations in n-particle gluon emission are generated by event averaging over color sources. Because the range of color correlations in the transverse plane is of order $1/\qs$, where $\qs$ is the saturation scale, our result suggests an intuitive picture of coherent multiparticle production arising from fluctuations in particle production from color flux tubes of size $1/\qs$.

These color flux tubes, called Glasma flux tubes~\cite{Lappi:2006fp}, are approximately boost invariant in rapidity because of the underlying boost invariance of the 
classical fields. The Glasma flux tubes also carry 
topological charge~\cite{Kharzeev:2001ev}, which may result in observable metastable CP-violating domains~\cite{Kharzeev:2007jp}. Albeit eqs.~(\ref{eq:ngluon-LLog}) and ~(\ref{eq:AA}) describe initial particle production from the Glasma flux tubes, they do not describe the growth of instabilities or subsequent final state interactions of the produced gluons which become important for times $\tau \geq 1/\qs$.  Issues related to the "unstable Glasma"  and its decay have been discussed by us elsewhere~\cite{Gelis:kn}. 

At large transverse momenta ($p_\perp \geq \qs$), analytical expressions for the single particle distributions in eq.~(\ref{eq:AA}) are available~\cite{Kovner:1995ts}. These were used to compute 2-gluon~\cite{Dumitru:2008wn}, 3-gluon~\cite{Dusling:2009ar} and n-gluon correlations~\cite{Gelis:2009wh} in the MV model~\cite{McLerran:1994ni}. One obtains~\cite{Gelis:2009wh}
 \begin{eqnarray}
\left< \frac{\ud^n N}{\ud y_1 \ud^2 \pt_1 \cdots \ud y_n \ud^2 \pt_n}\right> = 
\frac{(q-1)!}{k^{q-1}}  \left< \frac{\ud N}{\ud y_1 \ud^2 \pt_1}\right> 
\cdots 
\left< \frac{\ud N}{\ud y_n \ud^2 \pt_n }\right> \, ,
\label{eq:n-cumulant}
\end{eqnarray}
where  $k = \zeta_n (N_c^2 -1) \qs^2 S_\perp/2\,\pi$, $\zeta_n$ is a non--perturbative coefficient that can be determined from numerical simulations, and $\qs^2  S_\perp \sim N_{\rm part}$. The expression in Eq.~(\ref{eq:n-cumulant}) corresponds to a negative binomial distribution which is widely seen in hadronic scattering data. The variance in these distributions can be expressed as $\sigma^2 ={\bar {n^2}} -{\bar n}^2= {\bar n} + {\bar n}^2/k$. Both 
$\sigma^2$ and $k$ are proportional to $N_{\rm part}$, as seen in the RHIC data~\cite{Adare:2008ns}. This remarkably simple geometrical structure of multi-particle correlations was derived for $p_\perp \geq \qs$; however, there is now evidence from numerical lattice simulations of two particle correlations that this structure may be robust at low $p_\perp$ as well.

The ridge~\cite{Plenary} in the Glasma flux tube model is a combination of {\it initial state} long range rapidity correlations and {\it final state} radial flow which collimates the angular distribution of correlated pairs. From causality, the latest time at which gluon rapidity correlations could set in is given by the relation $\tau \leq \tau_{f} \exp(-0.5 \Delta y)$, where $\Delta y$ is the rapidity separation between pairs and 
$\tau_f$ is the freeze-out time. The persistence of the ridge to large $\Delta y$ suggests that rapidity correlations are formed at early times;  indeed they reflect correlations in the nuclear wavefunctions as suggested by eq.~(\ref{eq:ngluon-LLog}). The angular correlation can be seen already from simple radial boost models where the boost parameter is fitted to the single particle spectra~\cite{Dumitru:2008wn,Dusling:2009ar}.  A more sophisticated treatment gives a good fit to RHIC data on the centrality dependence and energy dependence of the amplitude of the ridge and of its azimuthal 
width~\cite{Gavin:2008ev}. Computations~\cite{Dusling:2009ar}. are also consistent with observed features of the data on three particle correlations.

The blast wave model of radial ``Hubble" flow needs to be improved to account for all global features of the ridge data. An example of this is the ``away side" ridge, which likely is due to global energy momentum conservation in each event. A first hydrodynamical calculation incorporating flux tube like structures in the initial conditions reproduces all such features of the data~\cite{Takahashi:2009ua}; however,  only ``triggered" ridge events have been considered thus far. With a realistic hydro model, ridge like correlations, when compared to data, could provide further insight into the development of radial flow and the persistence of long range correlations at late times~\cite{Shuryak:2009cy}.

\begin{figure}[htbp]
\begin{center}
\hfill
\resizebox*{6.5cm}{!}{\includegraphics{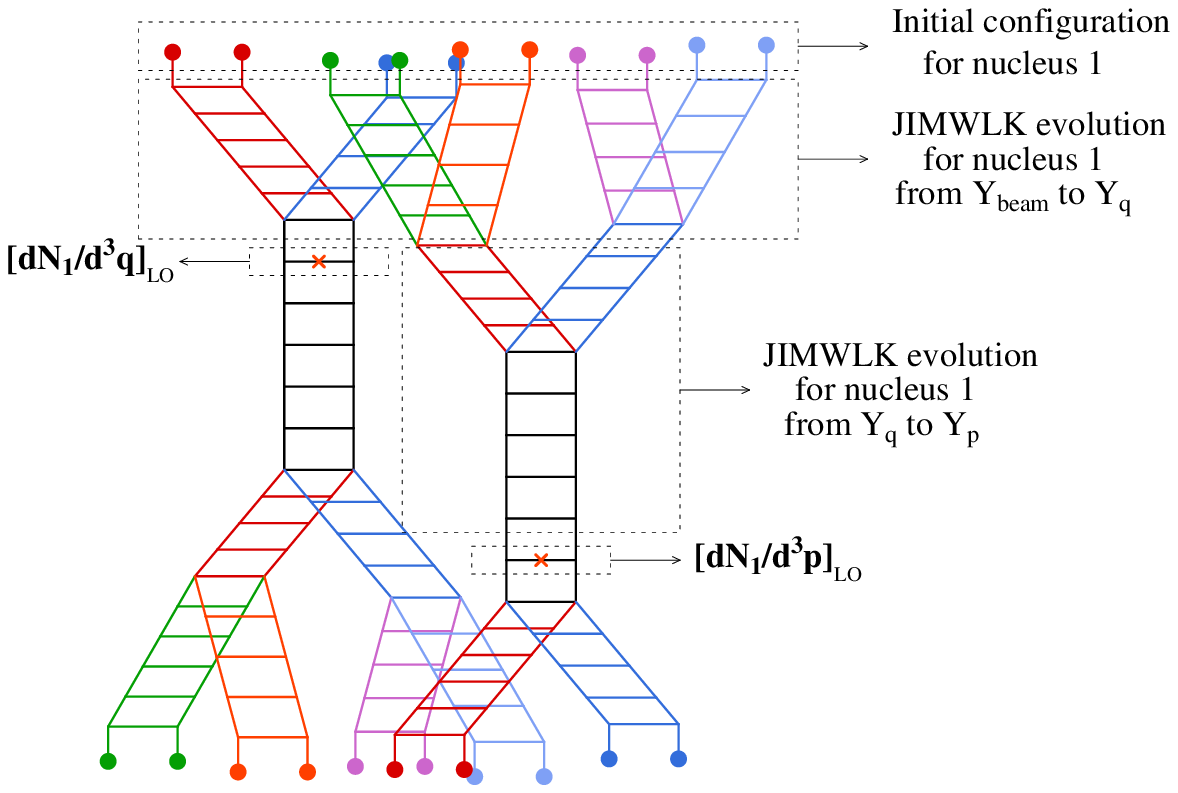}}
\hfill
\resizebox*{6.5cm}{!}{\includegraphics{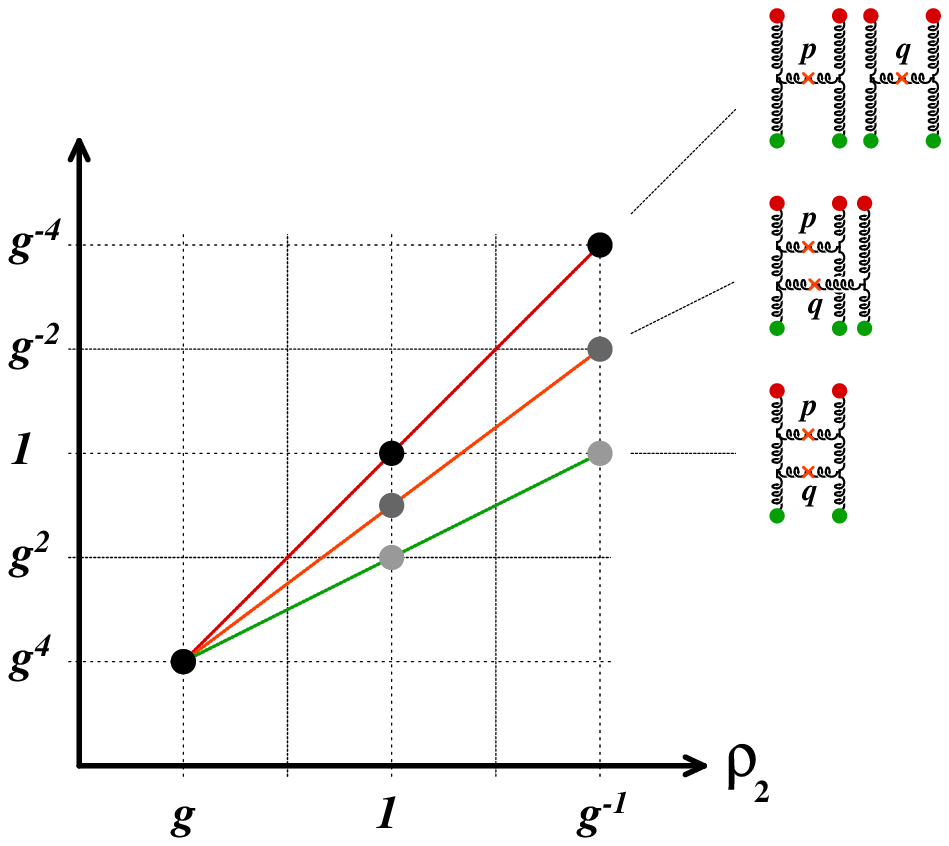}}
\hfill
\end{center}
\caption{\label{fig:QM}Left: Gluon ladder diagram demonstrating building blocks in QCD evolution of the inclusive
  2-gluon spectrum at large rapidity separations. The merging of conributions of several sources is typical of JIMWLK evolution. The crosses on rungs of the gluon ladders denote the positions in rapidity of the two tagged gluons. The evolution from nucleus 2 at the bottom of the figure is inferred. Right: Relative magnitude of leading order near-side Long Range Correlations (``Glasma")and 
  near-side Short Range (``pQCD") Correlations.  Color Charge density of projectile 1  ($\rho_1={\cal O}(g^{-1})$) is held fixed while that of target 2 ($\rho_2$) is varied. See text for details.}
\end{figure}

We now turn to genuine long range correlations. As stated, eq.~(\ref{eq:ngluon-LLog}) is valid for $\Delta y\leq 1/\as$, which corresponds approximately to $3$-$5$ units of rapidity at RHIC and the LHC respectively. The only energy dependence  in this expression comes in from the energy dependence of the weight functionals $Z$ (equivalently, $\qs$), while the dependence on $\Delta y$ is flat. This is an approximation because gluon radiation occurs in the rapidity interval between tagged gluons. Further, gluons are emitted in the presence of colored sources; the inclusive multiparticle spectrum at large $\Delta y$ is therefore sensitive to non-trivial multi-parton correlations in the nuclear wavefunctions. This problem has been solved~\cite{Gelis:2008sz}; the solution is represented diagrammatically in Fig.~(\ref{fig:QM}). The modification to eq.~(\ref{eq:ngluon-LLog}) is that one now has in addition Green functions describing the evolution from gluon $y_p$ to $y_q$ and vice-versa. These Green functions obey the JIMWLK equation. As the rapidities of the two gluons approach 
each other, one recovers eq.~(\ref{eq:ngluon-LLog}). Phenomenological work exploring the consequences of this novel formalism will be reported shortly.

We turn finally to discuss the relation of our work  on Glasma correlations  to correlations from perturbative QCD (``pQCD"). Examples of respective leading order diagrams and their power counting are shown in the right figure of Fig.~(\ref{fig:QM}). The Feynman diagram on top depicts two particle correlations from disconnected graphs in the Glasma. These generate the near-side long range rapidity correlations (LRC). The bottom figure depicts the contribution to two particle correlations from 2-to-2 Feynman graphs, analogous to conventional pQCD two particle correlations. These generate the near-side short range rapidity correlations (SRC). The middle graph is the contribution to 2 particle correlations from the interference between the two. When the parton densities in both nuclei are large (central collisions/high energies/low momenta), the cross-section for LRC dominates that of the SRC; as suggested by the positions of the filled dots, the disconnected graph contribution is larger than the interference graph by $1/\as$ and is larger than the pure SRC graph (at bottom) by as much as an order of magnitude ($1/\alpha_S^2$). In addition, one has a kinematic suppression of the latter cross-section. This is because ``pQCD" correlations are SRC for good reason-they die off rapidly with $\Delta y$. In contrast, as we have discussed here, the disconnected Glasma graphs are a natural QCD mechanism to generate long range correlations. SRC will begin to compete in magnitude with LRC as parton densities are decreased (peripheral collisions/pA/high momenta). In pA collisions, as schematically represented by the filled dots in 
the left hand corner of the plot, the two mechanisms are parametrically of the same order, even though one expects the relative kinematic suppression of pQCD to persist. In pp collisions, SRC dominate, except perhaps at very low momenta. These statements can be quantified for a unified (``the long and the short") QCD perspective on multi-particle correlations. Work on extending our formalism (and phenomenological consequences thereoff) to pA collisions is in progress.

\section*{Acknowledgments} % please check/modify, comment out or delete if not needed
F. G. is supported in part by Agence Nationale de la Recherche via
 the programme ANR-06-BLAN-0285-01. T. L. is supported by the Academy of Finland, project 126604 and R.V. is supported 
 under DOE Contract No. DE-AC02-98CH10886.

 % do not change 

\begin{thebibliography}{00} % do not change 
   
\bibitem{Gelis:2008rw} F.~Gelis, T.~Lappi and R.~Venugopalan, Phys. Rev. {\bf D78}, 054019 (2008); {\it ibid.}, {\bf D78}, 054020 (2008).

\bibitem{Gelis:2006yv}F.~Gelis and R.~Venugopalan, Nucl. Phys. {\bf A776}, 135 (2006); {\it ibid.}, {\bf 779}, 177 (2006).

\bibitem{McLerran:1994ni}L.~D. McLerran and R.~Venugopalan, Phys. Rev. {\bf D49}, 2233 (1994); E.~Iancu and R.~Venugopalan,
arXiv:hep-ph/0303204.

\bibitem{JalilianMarian:1998cb}J.~Jalilian-Marian, A.~Kovner, A.~Leonidov and H.~Weigert,Phys. Rev. {\bf D59}, 014014 (1998);
%%CITATION = HEP-PH 9706377;%%
E.~Iancu, A.~Leonidov and L.~D. McLerran, Nucl. Phys. {\bf A692}, 583 (2001).

\bibitem{Balitsky:1998kc}I.~Balitsky,Nucl. Phys. {\bf B463}, 99 (1996); Y.~V. Kovchegov, Phys. Rev. {\bf D60}, 034008 (1999).

\bibitem{Krasnitz:2003jw}A.~Krasnitz, Y.~Nara and R.~Venugopalan, Nucl.\ Phys.\  A {\bf 727}, 427 (2003); T.~Lappi,Phys.\ Rev.\  C {\bf 67}, 054903 (2003).

\bibitem{Lappi:2006fp}T.~Lappi and L.~McLerran, Nucl. Phys. {\bf A772}, 200 (2006).

\bibitem{Kharzeev:2001ev}D.~Kharzeev, A.~Krasnitz and R.~Venugopalan, Phys. Lett. {\bf B545}, 298 (2002).

\bibitem{Kharzeev:2007jp}D.~E.~Kharzeev, L.~D.~McLerran and H.~J.~Warringa, Nucl.\ Phys.\  A {\bf 803}, 227 (2008).

\bibitem{Gelis:kn}F.~Gelis, S.~Jeon and R.~Venugopalan, Nucl.\ Phys.\  A {\bf 817}, 61 (2009); F.~Gelis, T.~Lappi and R.~Venugopalan, Int.\ J.\ Mod.\ Phys.\  E {\bf 16}, 2595 (2007); K. Fukushima, F. Gelis and L. McLerran,  Nucl.\ Phys.\  A {\bf 786}, 107 (2007); P.~Romatschke and R.~Venugopalan, Phys.\ Rev.\ Lett.\  {\bf 96}, 062302 (2006).

\bibitem{Kovner:1995ts}A.~Kovner, L.~D. McLerran and H.~Weigert, Phys. Rev. {\bf D52}, 3809 (1995).

\bibitem{Dumitru:2008wn}A.~Dumitru, F.~Gelis, L.~McLerran and R.~Venugopalan, Nucl. Phys. {\bf A810}, 91 (2008).

\bibitem{Dusling:2009ar}K.~Dusling, D.~Fernandez-Fraile and R.~Venugopalan,  arXiv:0902.4435 [nucl-th].

\bibitem{Gelis:2009wh}F.~Gelis, T.~Lappi and L.~McLerran,arXiv:0905.3234 [hep-ph].

\bibitem{Adare:2008ns} A.~Adare {\em et~al.} [PHENIX], Phys. Rev. {\bf C78}, 044902 (2008).

\bibitem{Plenary}J.Putschke [STAR], these proceedings; W. Busza [PHOBOS], these proceedings.

\bibitem{Gavin:2008ev}S.~Gavin, L.~McLerran and G.~Moschelli, arXiv:0806.4718 [nucl-th].

\bibitem{Takahashi:2009ua}J.~Takahashi et al., arXiv:0902.4870 [nucl-th].

\bibitem{Shuryak:2009cy}E.~Shuryak,
  %``The Fate of the Initial State Perturbations in Heavy Ion Collisions,''
  arXiv:0903.3734 [nucl-th]; S.~J.~Lindenbaum and R.~S.~Longacre, arXiv:0809.2286 [nucl-th].

\bibitem{Gelis:2008sz}F.~Gelis, T.~Lappi and R.~Venugopalan, Phys. Rev. {\bf D79}, 094017 (2008), arXiv:0810.4829 [hep-ph].




\end{thebibliography}
\end{document}